\documentclass[twocolumn,amsmath,amssymb]{snp}
\usepackage{graphicx}
\usepackage{dcolumn}
\usepackage{bm}
\def\be{\begin{equation}}
\def\ee{\end{equation}}
\def\ba{\begin{array}}
\def\ea{\end{array}}
\def\bea{\begin{eqnarray}}
\def\eea{\end{eqnarray}}
\begin{document}

\title{{\Large Role of non-coplanarity in nuclear reactions using the Wong formula based on the proximity potential}}

\author{\large Manie Bansal}
\author{\large Raj K. Gupta}
\affiliation{Physics Department, Panjab University,Chandigarh-160014, INDIA}
\maketitle

\noindent {\large{\bf Introduction}}

Recently \cite{kumar09}, we assessed Wong's formula \cite{wong73} for its angular momentum $\ell$-summation and ``barrier 
modification" effects at sub-barrier energies in the dominant fusion-evaporation and capture (equivalently, quasi-fission) 
reaction cross-sections. For use of the multipole deformations (up to $\beta_4$) and (in-plane, $\Phi$=0$^0$) 
orientations-dependent proximity potential in fusion-evaporation cross-sections of $^{58}$Ni+$^{58}$Ni, $^{64}$Ni+$^{64}$Ni 
and $^{100}$Mo, known for fusion hindrance phenomenon in coupled-channels calculations, and the capture cross-sections of 
$^{48}$Ca+$^{238}$U, $^{244}$Pu and $^{248}$Cm reactions, forming superheavy nuclei, though the simple $\ell$=0 
barrier-based Wong formula is found inadequate, its extended version, the $\ell$-summed Wong expression fits very well the 
above noted capture cross-sections at all center-of-mass energies $E_{c.m.}$'s, but require (additional) modifications of 
the barriers to fit the fusion-evaporation cross-sections in the Ni-based reactions at below-barrier energies. Some barrier 
modification effects are shown \cite{kumar09} to be already present in Wong expression due to its inbuilt $\ell$-dependence 
via $\ell$-summation. 

In this paper, we study for the first time the dynamics of fission reactions, such as $^{11}$B+$^{235}$U and 
$^{14}$N+$^{232}$Th forming $^{246}$Bk$^*$ \cite{Behera01}, on the basis of the extended, $\ell$-summed Wong formula, 
including also the non-coplanarity ($\Phi\neq 0^0$) degree-of-freedom for all the three types of reactions, the 
fusion-evaporation, capture and fission cross-sections.

\noindent {\large{\bf The extended Wong model}}

Wong's expression for fusion cross-section due to colliding two deformed and oriented nuclei (orientations $\theta_i$), 
lying in two different planes (azimuthal angle $\Phi$ between the planes), in terms of $\ell$ partial waves, is 
\be 
\sigma (E_{c.m.},\theta_i,\Phi)=\frac{\pi}{k^2}\sum_{\ell=0}^{\ell_{max}}(2\ell+1)P_{\ell}(E_{c.m.},\theta_i,\Phi),
\label{eq:1} 
\ee
with $k=\sqrt{\frac{2\mu E_{c.m.}}{\hbar^2}}$, and $\mu$, the reduced mass. $P_{\ell}$ is the transmission coefficient for 
each $\ell$, describing, in Hill-Wheeler approximation, the penetration of barrier $V_{\ell}(R,E_{c.m.},\theta_i,\Phi)$.

Instead of solving Eq. (\ref{eq:1}) explicitly, which require the complete $\ell$-dependent potentials 
$V_{\ell}(R,E_{c.m.},\theta_i,\Phi)$, Wong summed it up {\it approximately}, using only $\ell$=0 quantities, which on 
replacing the $\ell$-summation in (\ref{eq:1}) by an integral, gives the Wong formula \cite{wong73}
\begin{small}
\bea
&&\sigma(E_{c.m.},\theta_i,\Phi)\nonumber \\
&=&\frac{{R_B^{0}}^{2}\hbar\omega_{0}}{2E_{c.m.}}
\ln\left[1+\exp\left(\frac{2\pi}{\hbar\omega_{0}}(E_{c.m.}-V_B^{0})
\right)\right].\nonumber\\
\label{eq:2}
\eea
\end{small}
\vspace{-0.5cm}
\par\noindent
Integrating (\ref{eq:2}) over $\theta_i$ and $\Phi$, we get the fusion cross-section $\sigma(E_{c.m.})$.

\begin{figure*}
\vspace{-1.0cm}
\includegraphics[width=2.2\columnwidth]{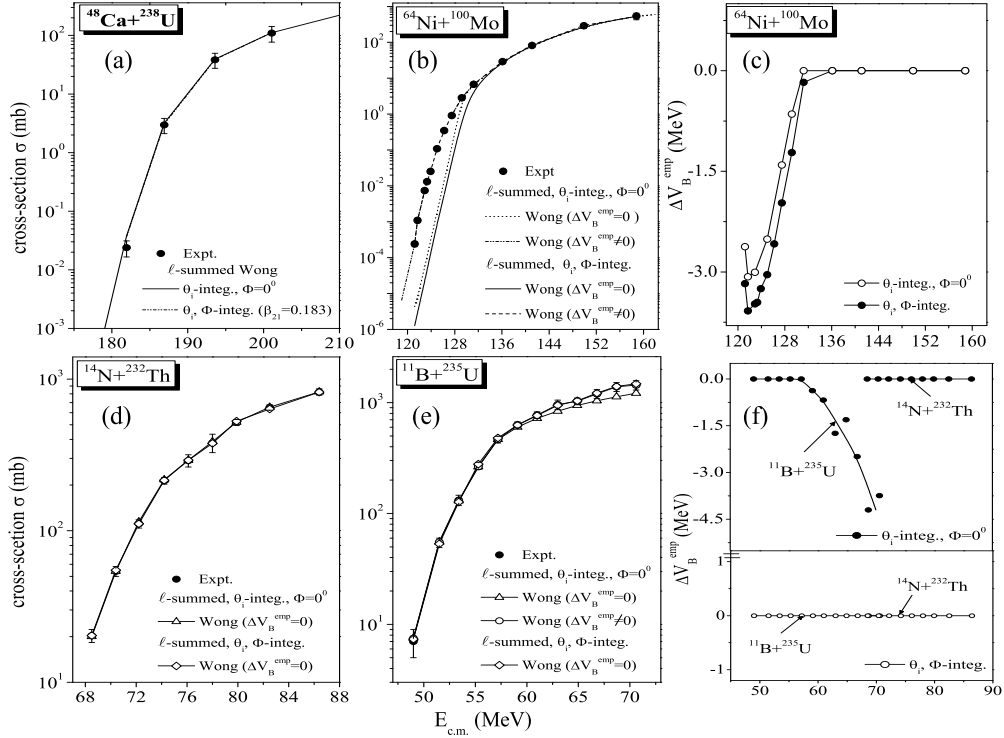}
\vspace{-1.3cm} 
\caption{$\ell$-summed Wong results for both $\Phi$=0$^0$ and $\Phi\ne$0$^0$, compared with experimental data, for: 
(a) $^{48}$Ca+$^{238}$U (b) and (c) $^{64}$Ni+$^{100}$Mo, and (d) to (f) $^{246}$Bk$^*$ due to $^{14}$N+$^{232}$Th and   
$^{11}$B+$^{235}$U channels.
}
\label{Fig.1}
\end{figure*}

For an explicit summation over $\ell$ in Eq. (\ref{eq:1}), the $\ell$-dependent interaction potential $V_{\ell}(R)$ is a
sum of Coloumb and nuclear proximity and centrifugal potentials, as
\bea
V_{\ell}(R)&=&V_P(R,A_{i},\beta_{\lambda i},T,\theta_{i},\Phi)+\frac{\hbar^2\ell(\ell+1)}{2\mu R^2}\nonumber\\
&&+V_{C}(R,Z_{i},\beta_{\lambda i},T,\theta_{i},\Phi),
\label{eq:3} 
\eea 
where, the $\ell$-summation in Eq. (\ref{eq:1}) is then carried out for the $\ell_{max}$ determined empirically for a best 
fit to measured cross-section. This procedure of explicit $\ell$-summation works very well for $\Phi$=0$^0$ case 
\cite{kumar09} in capture reactions $^{48}$Ca+$^{238}$U, $^{244}$Pu and $^{248}$Cm, but require further modification of 
the barrier for Ni-based reactions at sub-barrier energies, which could be carry out empirically \cite{kumar09} by either 
(i) keeping the curvature $\hbar\omega_{\ell}$ same and modifying the barrier height $V_B^{\ell}$, as
$$V_B^{\ell}(modified)=V_B^{\ell}+\Delta V_B^{emp},$$
or (ii) keep the barrier height $V_B^{\ell}$ same and modify the curvature $\hbar\omega_{\ell}$. We use here the method of 
modifying the barrier height.

\vspace{0.1cm}
\noindent {\large{\bf Calculations and results}}
\vspace{0.1cm}

The results of $\ell$-summed Wong expression (\ref{eq:1}) for both the cases of $\Phi$=0$^0$ and $\Phi\ne$0$^0$ are given 
in Fig. 1 for all the three types of reactions. Fig. 1(a) shows that the capture cross-section in $^{48}$Ca+$^{238}$U is 
fitted nicely even after giving a small deformation ($\beta_{21}$=0.183) to $^{48}$Ca for carrying out $\Phi$-integration. The
fitted $\ell_{max}(E_{c.m.})$ increase by one-to-two units. Similarly, the $^{64}$Ni+$^{100}$Mo reaction is fitted for both 
$\Phi$=0$^0$ and $\Phi\ne$0$^0$ by allowing empirically a small increase in ``barrier lowering" $\Delta V_B^{emp}$ 
(Figs. 1(b) and 1(c)). On the other hand, there is a strong entrance channel dependence in the case of fission reaction: 
whereas a nice fit is obtained for both $\Phi$=0$^0$ and $\Phi\ne$0$^0$ cases in $^{14}$N+$^{232}$Th channel (Fig. 1(d)), 
a large disagreement in cross-sections at higher energies (Fig. 1(e)) and hence a large ``barrier lowering" 
$\Delta V_B^{emp}$ (Fig. 1(f)) is obtained for $\Phi$=0$^0$ in $^{11}$B+$^{235}$U channel, which reduces to zero for 
$\Phi\ne$0$^0$ case. In other words, for fission of $^{246}$Bk$^*$, the inclusion of non-coplanarity gives a complete fit 
to data for both the reaction channels, without introducing $\Delta V_B^{emp}$. 



\end{document}